# Lattice Problems, Gauge Functions and Parameterized Algorithms

V. Arvind and Pushkar S. Joglekar

Institute of Mathematical Sciences
C.I.T Campus,Chennai 600 113, India
{arvind,pushkar}@imsc.res.in

**Abstract.** Given a $k$-dimensional subspace $M \subseteq \mathbb{R}^n$ and a full rank integer lattice $\mathcal{L} \subseteq \mathbb{R}^n$, the *subspace avoiding problem* SAP, defined by Blömer and Naewe [BN07], is to find a shortest vector in $\mathcal{L} \setminus M$. Treating $k$ as a parameter (in the sense of parameterized complexity), we obtain new parameterized approximation and exact algorithms for SAP based on the AKS sieving technique [AKS01].
  - Our first result is a randomized $(1+\epsilon)$-approximation algorithm for parameterized SAP that runs in time $2^{O(n)} \cdot (1/\epsilon)^k$, where the parameter $k$ is the dimension of the subspace $M$. Thus, we obtain a $2^{O(n)}$ time algorithm for $\epsilon = 2^{-O(n/k)}$.
  - Several of our algorithms work for all *gauge* functions as metric with some natural restrictions, in particular for all $\ell_p$ norms. We also prove an $\Omega(2^n)$ lower bound on the query complexity of AKS sieving based exact algorithms for SVP that accesses the gauge function as oracle.
  - Next, we give a $2^{O(n+k \log k)}$ exact algorithm for the parameterized SAP for any $\ell_p$ norm.

This implies a $2^{O(n)}$ time randomized algorithm for computing the $i^{th}$ successive minima of rank $n$ lattice for any $\ell_p$ norm if $i$ is $O(n/\log n)$. It is known that computing all $n$ successive minima's is equivalent to the problems CVP, SIVP [M08]. So our result can be thought of as a step forward in getting $2^{O(n)}$ time randomized algorithm for CVP. We also give a randomized $2^{O(n)}$ time algorithm for CVP if the input instance satisfies certain promise. We also give a new algorithm for the Theta-series problem for which parameterized hardness results are shown in [DFVW99]. Furthermore, we study a new parameterized version of SVP, CVP, and SAP and show that these parameterized CVP and SAP have randomized $s^{O(k)}$ time algorithms, where $k$ is the parameter and $s$ is the input size, and are hard for the class W[1].

## 1 Introduction

Fundamental algorithmic problems concerning integer lattices are the shortest vector problem (SVP) and the closest vector problem(CVP). Given a lattice $\mathcal{L} \subset \mathbb{R}^n$ by a basis, the shortest vector problem (SVP) is to find a shortest non zero vector in $\mathcal{L}$ w.r.t. some metric given by a *gauge* function in general (usually the $\ell_p$ norm for some $p$). Likewise, the closest vector problem (CVP) takes as input a lattice $\mathcal{L} \subset \mathbb{R}^n$ and vector $v \in \mathbb{R}^n$ and asks for a $u \in \mathcal{L}$ closest to $v$ w.r.t. a given metric. These problems have polynomial-time approximation algorithms based on the celebrated LLL algorithm for basis reduction [LLL82].

More recently, Ajtai, Kumar and Sivakumar in a seminal paper [AKS01], gave a $2^{O(n)} \cdot \text{poly}(s)$ time randomized algorithm to find a shortest nonzero vector in the lattice w.r.t. the $\ell_2$ norm, where $s$ is the size of the input in binary encoding. In subsequent work [AKS02] they also gave a $1 + \epsilon$ randomized approximation algorithm of similar running time for CVP. Their algorithms are based on a generic sieving procedure that exploits the underlying geometry.

Earlier, Kannan [Kan87] has given $2^{O(n \log n)} \cdot \text{poly}(s)$ time deterministic algorithms for exact solutions to both SVP and CVP. The best known deterministic algorithm for CVP is due to Blömer [Bl00] which runs in time $O(n! s^{O(1)})$ where $s$ is input size. Recently, Blömer and Naewe [BN07] introduced the *subspace avoiding* problem (SAP) that generalizes SVP and CVP. In SAP the input instance is a lattice $\mathcal{L}$ and subspace $M \subset \mathbb{R}^n$ and the goal is to find a shortest vector in $\mathcal{L} \setminus M$. Using a variant of the Ajtai-Kumar-Sivakumar sieving procedure Blömer and Naewe [BN07] describe a randomized $1 + \epsilon$ approximation algorithm for SAP that runs in time $(2 + 1/\epsilon)^n$, for any $\ell_p$ norm.

In the first part of the present paper we focus on SAP. We treat the dimension $\dim(M) = k$ as a parameter and obtain approximation and exact algorithms for SAP whose running time and approximation guarantee are sensitive to the parameter $k$. Our results are also based on the AKS sieving procedure [AKS01,Re]. We summarize our main results below:

- We give $2^{O(n+k\log 1/\epsilon)}$ time randomized algorithm to solve $1 + \epsilon$ approximation of the parameterized SAP problem, where $n$ is rank of the lattice and $k$ is the dimension of subspace. Our algorithm works w.r.t. any gauge function satisfying some natural conditions. This generalizes and improves on the result of Blömer and Naewe [BN07], when the dimension of the subspace involved is small. We also propose a generalization of SAP (the *convex body avoiding* problem) and give a singly exponential algorithm for $1 + \epsilon$ approximation of the problem. We use this to give an algorithm for the theta-series problem [DFVW99] for lattices.
- We give a randomized algorithm to exactly solve parameterized SAP with running time $2^{O(n+k\log k)}$. As an implication of this we obtain a $2^{O(n)}$ randomized algorithm to find $i^{th}$ successive minima for any $\ell_p$ norm if $i$ is $O(n/\log n)$. This also yields a $2^{O(n)}$ randomized algorithm that exactly solves CVP instances $(\mathcal{L}, v)$ with $d(\mathcal{L}, v) < \sqrt{3}/2 \cdot \lambda_t(\mathcal{L})$, where $\lambda_t(\mathcal{L})$ is $t^{th}$ successive minima of $\mathcal{L}$ w.r.t. $\ell_2$ norm for $t = O(n/\log n)$.
- Motivated by the parameterized complexity classes [DF99,DFVW99] we introduce and study a new parameterization of SVP, CVP, and SAP and show some upper and lower bound results.

Finally, we consider the unique-CVP problem. R. Kumar and D. Sivakumar in [KS99] have proved that unique-SVP is NP-hard under randomized reductions for $\ell_2$ norm. We give a randomized polynomial-time reduction from the search version of CVP to the search version of unique-CVP for any $l_p$ norm, $1 \leq p < \infty$. Likewise, we can give a similar reduction showing hardness of unique-SAP.

## 2 Preliminaries

A lattice $\mathcal{L}$ is a discrete additive subgroup of $R^n$, $n$ is called dimension of the lattice. For algorithmic purposes we can assume that $\mathcal{L} \subseteq \mathbb{Q}^n$, and even in some cases $\mathcal{L} \subseteq \mathbb{Z}^n$. A lattice is usually specified by a basis $B = \{b_1, \cdots, b_m\}$, where $b_i \in \mathbb{Q}^n$ and $b_i$'s are linearly independent. $m$ is called the rank of the lattice. The lattice is called full-rank if $m = n$. Though most of the results in the paper are applicable for general lattices, for convenience we mainly consider only full-rank lattices in this paper. For $x \in \mathbb{Q}^n$ let size($x$) denote the number of bits for representing $x$ in standard binary representation. Let size($\mathcal{L}$) denote number of bits for representing the basis vectors $b'_is$. Next we define a gauge function which generalizes the notion of usual $l_p$ norms.

**Definition 1 (gauge function [Si45]).** *A function $f : \mathbb{R}^n \to \mathbb{R}$ is called a* gauge function *if it satisfies following properties:*

1. $f(x) > 0$ *for all* $x \in \mathbb{R}^n \setminus \{0\}$ *and* $f(x) = 0$ *if* $x = 0$.
2. $f(\lambda x) = \lambda f(x)$ *for all* $x \in \mathbb{R}^n$ *and* $\lambda \in \mathbb{R}$.
3. $f(x + y) \leq f(x) + f(y)$ *for all* $x, y \in \mathbb{R}^n$.

For $v \in \mathbb{R}^n$ we denote $f(v)$ by $\|v\|_f$ and call it norm of $v$ with respect to gauge function $f$. It is easy to see that any $l_p$ norm satisfies all the above properties. Using a gauge function $f$ we can easily define a metric $d$ on $\mathbb{R}^n$ simply by assigning $d(x, y) = f(x - y)$ for all $x, y \in \mathbb{R}^n$. For $x \in \mathbb{R}^n$ and $r > 0$, let $B_f(x, r)$ denote a ball of radius $r$ around $x$ with respect to the gauge function $f$, i.e. $B_f(x, r) = \{y \in \mathbb{R}^n | f(x - y) \leq r\}$. We call $B_f(x, r)$ as f-ball of radius $r$ with center $x$. We denote the metric balls with respect to usual $l_p$ norm by $B_p(x,r)$. Unless specified particularly we are always considering balls in $n$ dimensional space. Next proposition characterizes the class of all gauge functions. We state it without proof. For the proof see, e.g. [Si45].

**Proposition 1.** *Let $f : \mathbb{R}^n \to \mathbb{R}$ be any gauge function then a unit radius ball around origin with respect to $f$ is a $n$ dimensional bounded O-symmetric convex body. Conversely for any $n$ dimensional bounded O-symmetric convex body $C$, there is a gauge function $f : \mathbb{R}^n \to \mathbb{R}$ such that $B_f(0,1) = C$.*

Given a ball of radius $r$ around origin with respect to a gauge function $f$, from the Proposition 1 it follows that $B_f(0,r)$ is an O-symmetric convex body. It is easy to check that for any $r > 0$ and any constant $c$ we have $\text{vol}(B_f(0,cr)) = c^n \text{vol}(B_f(0,r))$, where $\text{vol}(C)$ denotes the volume of the corresponding convex body $C$ (see e.g. [Si45]).

We make a technical assumption regarding a gauge function $f$. For some polynomial $p(n)$, $B_2(0, 2^{-p(n)}) \subseteq B_f(0,1) \subseteq B_2(0, 2^{p(n)})$, i.e. there exists a Euclidean sphere of radius $2^{-p(n)}$ inside the convex body $B_f(0,1)$, and $B_f(0,1)$ is contained inside a Euclidean sphere of radius $2^{p(n)}$. We call a gauge function given by a membership oracle as *nice gauge function* if it satisfies this property. Note that if $f$ is a nice gauge function and $v \in \mathbb{Q}^n$ we have size($f(v)$)=poly(n,size($v$)). For a nice gauge function $f$ we can sample points from convex body $B_f(0,r)$ almost uniformly at random in poly(size($r$),n) time using Dyer-Frieze-Kannan algorithm [DFK91]. Note that, this assumption is not restrictive and fairly standard, e.g. all $l_p$ norms $p \geq 1$ satisfy this property. The $i^{th}$ successive minima of a lattice $\mathcal{L}$ with respect to a gauge function $f$ is smallest $r > 0$ such that $B_f(0,r)$ contains at least $i$ linearly independent lattice vectors. It is denoted by $\lambda_i^f(\mathcal{L})$. We denote $i^{th}$ successive minima with respect to usual $l_p$ norm by $\lambda_i^p(\mathcal{L})$.

*Remark 1.* It is important to note that given an input lattice $\mathcal{L} \subset \mathbb{R}^n$ and a gauge function $f$ as an oracle, we have a linear transform $A : \mathbb{Z}^n \to \mathcal{L}$ where $\mathbb{Z}^n$ is the standard lattice, which is a lattice isomorphism. Furthermore, the function $g(x) = f(A(x)$ is a nice gauge function on $\mathbb{Z}^n$ with which we can work instead of $(\mathcal{L}, f)$. However, we find it convenient to allow for both $\mathcal{L}$ and $f$ to be input.

## 3   Approximation algorithm for SAP for any gauge function

We first describe the AKS sieving procedure [AKS01] for any gauge function, analyze its running time and explain its key properties. The following lemma is crucially used in the algorithm.

**Lemma 1.** *[Sieving Procedure] Let $f : \mathbb{R}^n \to \mathbb{R}$ be any gauge function. Then there is a sieving procedure that takes as input a finite set of points $\{\mathbf{v}_1, \mathbf{v}_2, \mathbf{v}_3, \ldots, \mathbf{v}_N\} \subseteq B_f(0,r)$, and in $N^{O(1)}$ time it outputs a subset of indices $S \subset [N]$ such that $|S| \leq 5^n$ and for each $i \in [N]$ there is a $j \in S$ with $f(\mathbf{v}_i - \mathbf{v}_j) \leq r/2$.*

*Proof.* The sieving procedure is exactly as described in O. Regev's lecture notes [Re]. The sieving procedure is based on a simple greedy strategy. We start with $S = \emptyset$ and run the following step for all elements $v_i, 1 \leq i \leq N$. At the $i^{\text{th}}$ step we consider $v_i$. If $f(v_i - v_j) > r/2$ for all $j \in S$ include $i$ in the set $S$ and increment $i$. After completion, for all $i \in [N]$ there is a $j \in S$ such that $f(v_i - v_j) \leq r/2$. The bound on $|S|$ follows from a packing argument combined with the fact that $\text{vol}(B_f(0,cr)) = c^n \text{vol}(B_f(0,r))$ for any $r > 0$ and a constant $c > 0$. More precisely, for any two points $v_i, v_j \in S$ we have $f(v_i - v_j) > r/2$. Thus, all the convex bodies $B_f(v_i, r/4)$ for $v_i \in S$ are mutually disjoint and are contained in $B_f(0, r + r/4)$. Also note that $\text{vol}(B_f(0, dr)) =$ $d^n \text{vol}(B_f(0,r))$ for any constant $d > 0$. It follows that $5^n \text{vol}(B_f(v_i, r/4)) \geq \text{vol}(B_f(0, r + r/4))$. Hence, $|S| \leq 5^n$. The second property of $S$ is guaranteed by the sieving procedure. ∎

We give a $2^{O(n)} \cdot (1/\epsilon)^{O(k)}$ time randomized algorithm for computing a $(1 + \epsilon)$ approximate solution to an instance $(\mathcal{L}, M)$ of SAP, treating the dimension $\dim(M) = k$ as a parameter. This algorithm is based on a different analysis of the AKS sampling procedure than described by Blömer and Naewe [BN07], and it works for any gauge function. Suppose, without loss of generality, that the input lattice $\mathcal{L} \subseteq \mathbb{R}^n$ is $n$-dimensional given by a basis $\{b_1, \cdots, b_n\}$, so that $\mathcal{L} = \sum_{i=1}^{n} \mathbb{Z} \cdot b_i$. Let us fix a nice gauge function $f$ and let $v \in \mathcal{L}$ denote a shortest vector in $\mathcal{L} \setminus M$, i.e. $f(x)$ for $x \in \mathcal{L} \setminus M$ attains minimum value at $x = v$. Let $s = \text{size}(\mathcal{L}, M)$ denote the input size (which is the number of bits for representing the vectors $b_i$ and the basis for $M$). As $v$ is a shortest vector in $\mathcal{L} \setminus M$ and $f$ is a nice gauge function it is quite easy to see that $size(f(v))$ is bounded by a polynomial in $s$. Thus, we can scale the lattice $\mathcal{L}$ to ensure that $2 \leq f(v) \leq 3$. More precisely, we can compute polynomially many scaled lattices from $\mathcal{L}$, so that $2 \leq f(v) \leq 3$ holds for at least one scaled lattice. Thus, we can assume that $2 \leq f(v) \leq 3$ holds for the lattice $\mathcal{L}$.

**Proposition 2.** *Let $\mathcal{L} \subset \mathbb{R}^n$ be a rank $n$ lattice, $v \in \mathcal{L}$ such that $2 \leq f(v) \leq 3$ for a nice gauge function $f$. Consider the convex regions $C = B_f(-v, 2) \cap B_f(0, 2)$ and $C' = B_f(v, 2) \cap B_f(0, 2)$. Then $C' = C + v$ and $vol(C) = vol(C') = \Omega(\frac{vol(B_f(0,2))}{2^{O(n)}})$.*

The Proposition 2 immediately follows since $B_f(-v/2, 1/2) \subseteq C, B_f(v/2, 1/2) \subseteq C'$. Next, our algorithm follows the usual AKS random sampling procedure.

Let $R = n \cdot max_i \|b_i\|_f$. It is clear that $\text{size}(R)$ is polynomial in $s$ since $f$ is a nice gauge function. Let $B_f(0, 2)$ denote the $f$-ball of radius 2 around the origin. Since we have an oracle for membership in $B_f(0, 2)$ we can almost uniformly sample from $B_f(0, 2)$ using a well-known algorithm [DFK91]. Let $x_1, x_2, \cdots, x_N$ denote such a random sample, for $N = 2^{c \cdot (n + k \log(1/\epsilon))} \cdot \log R$ where the constant $c > 0$ will be suitably chosen. Now, using the lattice $\mathcal{L}$ we can round off the points $x_i$. More precisely, we express $x_i = \Sigma_j \alpha_{ij} b_j$ for rationals $\alpha_{ij}$. Then, from each vector $x_i$ we compute the vector $y_i = \Sigma_j \beta_{ij} b_j$, where $0 \leq \beta_{ij} < 1$, by adding appropriate integral multiples of the $b_j$'s to the expression for $x_i$. Thus, the points $y_1, \cdots, y_N$ are in the interior of the fundamental parallelepiped of $\mathcal{L}$, and each $x_i - y_i \in \mathcal{L}$. We denote this by $y_i = x_i(\text{mod } \mathcal{L})$. We now have the set of $N$ pairs $Z = \{(y_i, x_i) \mid i \in [N]\}$, where $x_i - y_i$ are lattice points. Since $y_i$ lie inside the fundamental parallelepiped we have $\|y_i\|_f \leq n \cdot max_i \|b_i\|_f = R$ for $i = 1$ to $N$.

Now, we apply the AKS sieving procedure in Lemma 1 to the set $\{y_1, y_2, \cdots, y_N\}$. The result is a subset $S \subset [N]$ of at most $5^n$ indices such that for each $i \in [N]$ there is some $j \in S$ such that $f(y_i - y_j) \leq R/2$. We remove from $Z$ all $(x_j, y_j)$ for $j \in S$ and replace each remaining $(x_i, y_i) \in Z$ by a corresponding $(x_i, y_i - (y_j - x_j))$, where $j \in S$ is the first index such that $f(y_i - y_j) \leq R/2$. After the sieving round, the set $Z$ has the property that for each $(x_i, z_i) \in Z$ we have $x_i - z_i \in \mathcal{L}$ and $f(x_i - z_i) \leq 2 + R/2$, and $Z$ has shrunk in size by at most $5^n$. We continue with $O(\log R)$ sieving rounds so that we are left with a set $Z$ with $N - O(\log R)5^n$ pairs $(x_i, z_i)$ such that $x_i - z_i \in \mathcal{L}$ and $f(x_i - z_i) \leq 8$. We can ensure that $|Z| \geq 2^{c_1(n + k \log(1/\epsilon))}$ for an arbitrary constant $c_1$ by appropriately choosing constant $c$. The vectors, $x_i - z_i$ for $(x_i, z_i) \in Z$ follows some distribution among lattice points inside $B_f(0, 8)$.

We now describe the approximation algorithm, its proof of approximation guarantee and running time bound in the following theorem.

**Theorem 1.** *Let $\mathcal{L}$ be a lattice with a target vector $v \in \mathcal{L} \setminus M$ such that $2 \leq f(v) \leq 3$ for a given gauge function $f$ and $f(v) \leq f(x)$ for all $x \in \mathcal{L} \setminus M$. Then there is a randomized algorithm that in* time $2^{O(n+k\log(1/\epsilon))}$.poly(size($\mathcal{L}$)) *computes a set $Z$ of pairs $(x_i, z_i)$ such that $|Z| \geq 2^{c_1 \cdot (n+k\log(1/\epsilon))}$ for a constant $c_1$ and $f(x_i - z_i) \leq 8$ for all $(x_i, z_i) \in Z$. Moreover, $z_i - x_i \in \mathcal{L}$ are such that with probability $1 - 2^{-O(n)}$ there is a pair of points $(x_i, z_i), (x_j, z_j) \in Z$ such that $v + u = (x_i - z_i) - (x_j - z_j)$ for a vector $u \in \mathcal{L}$ with $f(u) \leq \epsilon$.*

*Proof.* Consider the set $Z$ of pairs $(x_i, z_i)$, obtained after the AKS sieving as described above, such that $|Z| \geq 2^{c_1(n+k\log(1/\epsilon))}$, and $f(x_i - z_i) \leq 8$ for all $(x_i, z_i) \in Z$. Consider the $k$-dimensional sublattice $L' = \mathcal{L} \cap M$ whose basis we can efficiently compute from $\mathcal{L}$ and $M$. Write $Z$ as a partition $Z = \bigcup_{j=1}^m Z_j$, where for each $Z_j$ there is a distinct coset $L' + v_j$ of $L'$ in $\mathcal{L}$ such that $z_i - x_i \in L' + v_j$ for all $(x_i, z_i) \in Z_j$. Let $Z'_j = \{z_i - x_i \mid (x_i, z_i) \in Z_j\}$. Suppose $u_j \in Z'_j \subseteq L' + v_j$ for $j = 1$ to $m$. Then, as $f(v) \geq 2$, $f(u_j - u_\ell) \geq 2$ for $j \neq \ell$. Hence unit radius $f$-balls around $u_i$'s are disjoint. Since $\mathrm{vol}(B_f(0, 8))/\mathrm{vol}(B_f(0, 1)) \leq 2^{c'n}$ for some constant $c'$, we will choose the constant $c_1$ large enough so that $|Z_{i'}| \geq 2^{d'(n+k\log(1/\epsilon))}$ for some large constant $d'$ and some index $i'$. Let $W = \{z_i - x_i \mid (x_i, z_i) \in Z_{i'}\}$. Now, let $T$ be any subset of $W$ so that $f(\beta - \beta') > \epsilon$ for all $\beta, \beta' \in T$. Then $B_f(\beta, \epsilon/2)$ are all disjoint $k$-dimensional $f$-balls for $\beta \in T$. By a packing bound it follows that $|T| \leq \left(\frac{8+\epsilon/2}{\epsilon/2}\right)^k$ which is $2^{O(k\log(1/\epsilon))}$. Note that we can use packing argument for $k$-dimensional balls since $W$ is contained in a coset of $k$ dimensional subspace $M$. Thus, for suitably large constant $d'$ we will have $2^{O(n)}$ many $\beta \in W$ that are inside $B_f(\beta', \epsilon/2)$ for a specific $\beta' \in T$.

Now, we apply the argument as explained in Regev's notes [Re] to reason with a modified distribution of the $x_i$. Firstly, by Proposition 2 we have convex region $C = B_f(0, 2) \cap B_f(-v, 2) \subset B_f(0, 2)$ such that $\mathrm{vol}(C) \geq \mathrm{vol}(B_f(0, 2)) \cdot 2^{-O(n)}$, $C' = C + v \subset B_f(0, 2)$ and $C \cap C + v = \emptyset$. Since $\mathrm{vol}(C) \geq \mathrm{vol}(B_f(0, 2)) \cdot 2^{-O(n)}$ and $|Z| \geq 2^{c_1 \cdot (n+k\log(1/\epsilon))}$ where we can choose the constant $c_1$ suitably large, it follows that in the beginning with high probability there are $2^{O(n+k\log(1/\epsilon))}$ pairs $(x_i, z_i) \in Z$ such that $x_i \in C$. Now, notice that we can replace the original distribution of $x_i$ with a modified distribution in which we output $x_i$ if it lies in $B_f(0, 2) \setminus (C \cup C')$ and if $x_i \in C$ it outputs either $x_i$ or $x_i + v$ with probability $1/2$ each. Similarly, if $x_i \in C' = C + v$ it outputs either $x_i$ or $x_i - v$ with probability $1/2$ each. This modified distribution coincides with the original distribution, and is introduced only for the purpose of analysis. Now, putting it together with the previous argument, this is easily seen to imply that with high probability we are likely to see $v + (\beta' - \beta)$ as the difference of $z_i - x_i$ and $z_j - x_j$ for some two pairs $(x_i, z_i), (x_j, z_j) \in Z$, for some $\beta', \beta \in T$. Since $v + (\beta' - \beta) \in \mathcal{L} \setminus M$ and $f(\beta' - \beta) \leq \epsilon$, $v + (\beta' - \beta)$ is actually $1 + \epsilon$ approximation of $v$. The actual algorithm will examine all $(z_i - x_i) - (z_j - x_j)$ for $(x_i, z_i), (x_j, z_j) \in Z$ obtained after sieving and output that element in $\mathcal{L} \setminus M$ of minimum $f$-value. This completes the proof of correctness of the approximation algorithm. ∎

An immediate consequence of the above theorem is a randomized $1 + \epsilon$ approximation algorithm for the parameterized SAP problem that runs in time $2^{O(n+k\log\frac{1}{\epsilon})} \cdot poly(size(\mathcal{L}, M))$.

**Corollary 1.** *Given a rank $n$ lattice $\mathcal{L}$ and a $k$-dimensional subspace $M \subset \mathbb{R}^n$, there is $1 + \epsilon$ randomized approximation algorithm for parameterized SAP (for any nice gauge function) with running time $2^{O(n+k\log\frac{1}{\epsilon})} \cdot poly(size(\mathcal{L}, M))$.*

## 4  An exact $2^{O(n+k\log k)}$ algorithm for SAP

In this section we give an exact algorithm for SAP. In particular, given a basis $\{b_1, \cdots, b_n\}$ of an integer lattice $\mathcal{L} \subseteq \mathbb{R}^n$ of rank $n$ and a subspace $M \subseteq \mathbb{R}^n$ of dimension $k$ we give a randomized algorithm to find a shortest vector in $\mathcal{L} \setminus M$ with respect to $l_p$ norm in time $2^{O(n+k\log k)}poly(s)$ where $s = size(\mathcal{L}, M)$. As an immediate corollary we get a $2^{O(n)}$ randomized algorithm to find $i^{th}$ successive minima w.r.t. any $l_p$ norm in an integer lattice $\mathcal{L}$ if $i$ is $O(n/\log n)$. Using ideas similar to Kannan's reduction from CVP to SVP ([Kan87]) we also get a hierarchy of algorithms for CVP depending upon the distance of given point from the lattice. Our exact algorithm uses the same sieving procedure as an approximation algorithm for SAP described in the Section 3. Note that without loss of generality we can assume that a shortest vector $v \in \mathcal{L} \setminus M$ satisfies $2 \leq \|v\|_p \leq 3$. Next we describe the algorithm.

1. Let $N = 2^{cn} \log(n.max_i \|b_i\|_p)$ and pick $x_1, x_2, \cdots, x_N$ uniformly at random from $B_p(0, 2)$.
2. Let $y_i = x_i (\mod \mathcal{L})$ and apply the AKS sieving procedure on the set $\{(x_1, y_1), \cdots, (x_N, y_N)\}$ as described in the Section 3 until for all tuples $(x_i, z_i)$ left after the sieving, $\|x_i - z_i\|_p \leq 8$.
3. Let $Z = \{(x_i, z_i) | i \in T\}, T \subset [N]$ be the set of tuples left after the sieving procedure. For all $i, j \in T$ compute lattice points $v_{i,j} = (z_i - x_i) - (z_j - x_j)$.
4. Let $w_{i,j}$ is a closest lattice vector to $v_{i,j}$ in the rank $k$ lattice $P = \mathcal{L} \cap M$ and let $r_{i,j} = v_{i,j} - w_{i,j}$. Output a vector of least non zero $\ell_p$ norm among all the vectors $r_{i,j}$ for $i, j \in T$.

First we prove the correctness of the algorithm.

*Claim.* For appropriate choice of the constant $c$ the algorithm outputs a shortest non zero vector in $\mathcal{L} \setminus M$ with respect to $\ell_p$ norm.

*Proof.* Let $v$ is a shortest vector in $\mathcal{L} \setminus M$. Consider the set of tuples left after the sieving procedure $Z = \{(x_i, z_i) | i \in T\}, T \subset [N]$. From arguments in the Theorem 1 it follows that for appropriate choice of the constant $c$ with "good" probability $Z$ contains tuples $(x_i, z_i), (x_j, z_j)$ such that $x_i, x_j \in C = B_p(0, 2) \cap B_p(-v, 2)$ and $z_i - x_i, z_j - x_j$ lie in the same coset of $P$, i.e. if $\beta_i = z_i - x_i$ and $\beta_j = z_j - x_j$ then $\beta_i - \beta_j \in P$. In fact we argued that there are "many" such tuples, the fact which is not essential for the current theorem. Since $x_i$ is chosen uniformly at random from $B_p(0, 2)$ we can replace it by equivalent uniform random sample as in Theorem 1. Since $x_i \in B_p(0, 2) \cap B_p(-v, 2)$ we replace $x_i$ by $x_i + v$ with probability $1/2$. Therefore in the Step 3 of the algorithm with good probability we have $v_{i,j} = (z_i - x_i - v) - (z_j - x_j) = -v + \beta_i - \beta_j$. Let $w_{i,j} \in P$ is a closest vector to $v_{i,j}$. So we have $d(v_{i,j}, w_{i,j}) \leq d(v_{i,j}, \beta_i - \beta_j) = \| - v\|_p$, i.e. $\|v_{i,j} - w_{i,j}\|_p \leq \|v\|_p$. But since $v_{i,j} - w_{i,j} \notin P$ and $v$ is a shortest vector in $\mathcal{L} \setminus M$, this implies $\|v_{i,j} - w_{i,j}\|_p = \|v\|_p$. So with good probability in Step 4 the algorithm will output a vector $r_{i,j}$ with $\|r_{i,j}\|_p = \|v\|_p$. This proves the correctness of the algorithm. ∎

Next we show that the running time of the algorithm is $2^{O(n+k \log k)} \cdot poly(s)$ where $s$ is the input size. In Step 1 of the algorithm we are sampling $N$ points from $B_p(0, 2)$, a ball of radius 2 with respect to $l_p$ norm. Since $B_p(0, 2)$ is a convex body, the task can be accomplished using Dyer-Frieze-Kannan algorithm [DFK91] in time $2^{O(n)} \cdot poly(s)$. It easily follows that the sieving procedure in Step 2 can be performed in $2^{O(n)}$ time. Note that $P$ is a rank $k$ lattice and a basis for $P$ can be efficiently computed in polynomial time using linear algebra. In the Step 4 of the algorithm we are solving $2^{O(n)}$ many instances of CVP for the rank $k$ lattice $P$. For $i, j \in S$ a closest vector to $v_{i,j}$ in the rank $k$ lattice $P$ can be computed in $2^{O(k \log k)}$ time using Kannan's algorithm for CVP [Kan87]. Hence the Step 4 takes $2^{O(n+k \log k)}$ time. Therefore the overall running time of the algorithm is $2^{O(n+k \log k)} \cdot poly(s)$. Note that by repeating above algorithm $2^{O(n)}$ times we can make the success probability of the algorithm exponentially close to 1.

**Theorem 2.** *Given an integer lattice $\mathcal{L}$ of rank $n$ and a subspace $M \subseteq \mathbb{R}^n$ of dimension $k < n$, There is a randomized algorithm to finds $v \in \mathcal{L} \setminus M$ with least possible $l_p$ norm. The running time* of the algorithm is $2^{O(n+k \log k)}$ times a polynomial in the input size and it succeeds with probability $1 - 2^{-cn}$ for an arbitrary constant c.

Given an integer lattice $\mathcal{L}$, Blömer and Naewe [BN07] gave $2^{O(n)}$ time $1+\epsilon$ factor approximation algorithm to compute $i^{th}$ successive minima in the lattice $\mathcal{L}$ for $i = 1$ to $n$. In the recent paper Micciancio [M08] has shown that the problem of computing linearly independent vectors $v_1, \ldots, v_n \in \mathcal{L}$ such that $\|v_i\| \leq \lambda_i(\mathcal{L})$ is deterministic poly-time equivalent to the problems CVP and SIVP. As an easy implication of the Theorem 2 we get a $2^{O(n)}$ algorithm to compute $i^{th}$ successive minima for $i \leq cn/\log n$ for an arbitrary constant c.

**Corollary 2.** *Given an integer lattice $\mathcal{L}$ of rank n and a positive integer $i \leq n$, there is a randomized algorithm with running time $2^{O(n+i \log i)} \cdot poly(size(\mathcal{L}))$ to compute $v_i \in \mathcal{L}$ such that $\|v_i\|_p = \lambda_i^p(\mathcal{L})$. In particular if $i \leq cn/\log n$ for an arbitrary constant c then there is an exact $2^{O(n)} \cdot poly(size(\mathcal{L}))$ time randomized algorithm to compute $i^{th}$ successive minima of the lattice $\mathcal{L}$.*

**Corollary 3.** *Given integer lattice $\mathcal{L}$ of rank n and $v \in \mathbb{Q}^n$ with promise that $d(v, \mathcal{L}) < \sqrt{3}/2\lambda_t(\mathcal{L})$, $t \leq n$. Where $\lambda_t(\mathcal{L})$ denotes $t^{th}$ successive minima of $\mathcal{L}$ with respect to $\ell_2$ norm, then there is a $2^{O(n+t \log t)} \cdot poly(size(\mathcal{L}))$ time randomized algorithm to compute a closest lattice point to v.*

*Proof.* By Corollary 2 we first compute $\lambda_t(\mathcal{L})$. We now use ideas from Kannan's CVP to SVP reduction [Kan87]. Let $b_1, b_2, \cdots, b_n$ be a basis for $\mathcal{L}$. We obtain new vectors $c_i \in \mathbb{Q}^{n+1}$ for $i = 1$ to $n$ by letting $c_i^T = (b_i^T, 0)$. Likewise, define $u \in \mathbb{Q}^{n+1}$ as $u^T = (v^T, \lambda_t/2)$. Let $\mathcal{M}$ be the lattice generated by the $n + 1$ vectors $u, c_1, c_2, \cdots c_n$. Compute the vectors $v_j \in \mathcal{M}$ such that $\|v_j\|_2 = \lambda_j(\mathcal{M})$ for $j = 1$ to $t$ using Corollary 2 in time $2^{O(n+t \log t)} \cdot poly(size(\mathcal{L}))$. Write vectors $v_j$ as $v_j = u_j + \alpha_j u$, $u_j \in \mathcal{L}(c_1, \cdots, c_n)$ and $\alpha_j \in \mathbb{Z}$. Clearly, $|\alpha_j| \leq 1$ since $u$ has $\lambda_t/2$ as its $(n + 1)^{th}$ entry. As $d(v, \mathcal{L}) < \sqrt{3}/2\lambda_t(\mathcal{L})$ we have $d(u, \mathcal{M}) < \lambda_t(\mathcal{L})$. Hence, there is at least one index $i, 1 \leq i \leq t$ such that $|\alpha_i| = 1$. Consider the set $S = \{v_i \mid 1 \leq i \leq t, |\alpha_i| = 1\}$ and let $u_j$ be the shortest vector in S. Writing $u_j = (w_j^T, 0)$, it is clear that the vector $-w_j \in \mathcal{L}$ is closest vector to v if $\alpha_j = 1$ and $w_j$ is a closest vector to v if $\alpha_j = -1$. ∎

## 5 The Theta-Series Problem for lattices

The Theta-Series problem for integer lattices takes as input a lattice $\mathcal{L}$ and an integer parameter k and asks if $\mathcal{L}$ has a vector such that $\|v\|_2^2 = k$. This problem is shown to be hard for both NP and parameterized complexity class W[1] in the paper by Fellows et al [DFVW99]. However, they do not show any upper bound results; In this section we apply the AKS sieving method to give a $(kp)^{O(n)}$ algorithm for solving the Theta-Series problem for all $\ell_p$-norms. We actually solve a more general problem defined below.

**Definition 2. Convex-Body Avoiding Problem (CAP)** *For an integer lattice $\mathcal{L}$ of rank n and an O-symmetric convex body C in $\mathbb{R}^n$, the* convex-body avoiding problem *is to find a $v \in \mathcal{L} \setminus C$ with least possible $\ell_p$ norm, where we assume that the convex body C is given by a membership oracle.*

**Theorem 3.** *Given integer lattice $\mathcal{L}$ of rank n and an O-symmetric convex body C in $\mathbb{R}^n$, there is $1+\epsilon$ factor approximation algorithm to solve CAP (w.r.t. any $\ell_p$ norm) with running time $2^{O(n) \cdot \log(1/\epsilon)} \cdot poly(size(\mathcal{L}))$.*

*Proof.* We claim that it suffices to solve the problem for the case when $C$ is $n$-dimensional. To see this, suppose $C$ is contained in some $k$-dimensional subspace $M$ of $\mathbb{R}^n$. We can find a basis for $M$ with high probability by using the polynomial-time almost uniform sampling algorithm from $C$ using [DFK91]. Next, we compute the lattice $\mathcal{L} \cap M$ and find a $(1+\epsilon)$ approximate solution $u$ for the $k$-dimensional convex body avoidance for the lattice $\mathcal{L} \cap M$ and $C$. We also solve the SAP instance $(\mathcal{L}, M)$ and find a $(1+\epsilon)$ approximate solution $v \in \mathcal{L} \setminus M$ using Theorem 1. The smaller of $u$ and $v$ is clearly a $(1+\epsilon)$ approximate solution for the input CAP instance.

Thus, we suppose $C$ is $n$ dimensional. Let $v$ be a shortest vector in $\mathcal{L} \setminus C$ which, as before, we can assume satisfies $2 \leq \|v\|_p \leq 3$ by considering polynomially many scalings of the lattice and the convex body. As in Theorem 1, we pick random points $x_1, \cdots, x_N$ from $B_p(0,2)$ for $N = 2^{cn \log(1/\epsilon)} \cdot poly(s)$. The constant $c > 0$ will be suitably chosen later. Let $y_i = x_i (\bmod \mathcal{L})$ for $i = 1$ to $N$. We apply several rounds of the AKS sieving on the set $\{(x_1, y_1), \cdots, (x_N, y_N)\}$ until we are left with a set $S$ of $2^{c_1 n \log(1/\epsilon)}$ pairs $(x_i, z_i)$ such that $\|x_i - z_i\|_p \leq 8$. From proposition 2 it follows easily that with good probability we have $Z \subseteq S$ such that $|Z| \geq 2^{c_2 n \log(1/\epsilon)}$ and for all $(x_i, z_i) \in Z$ we have $x_i \in D \cup D'$ where $D = B_p(0,2) \cap B_p(-v, 2)$ and $D' = B_p(0,2) \cap B_p(v, 2)$. Note that the the constant $c_2$ can be chosen as large as we like by appropriate choice of $c$. Let $Z' = \{z_i - x_i \mid (x_i, z_i) \in Z\}$. Now consider $\ell_p$ ball of radius $\epsilon/2$ centered at each lattice point $\beta \in Z'$. It is clear that for all $\beta \in Z'$, $B_p(\beta, \epsilon/2) \subseteq B_p(0, 8 + \epsilon/2)$. If for all $\beta \in Z'$ $\ell_p$ balls $B_p(\beta, \epsilon/2)$ are mutually disjoint, by packing argument we get $|Z'| \leq \frac{(8+\epsilon/2)^n}{(\epsilon/2)^n} = 2^{c'n \log(1/\epsilon)}$ for a constant $c'$. We choose constant $c$ appropriately to ensure that $c_2 > c'$. This implies that there exists tuples $(x_i, z_i), (x_j, z_j) \in Z$ such that $\|\beta_i - \beta_j\| \leq \epsilon$, where $\beta_i = z_i - x_i$ and $\beta_j = z_j - x_j$. Let $\beta = \beta_i - \beta_j$. We claim that it is not possible that both $\beta + v, \beta - v$ lie inside the convex body $C$. Because this implies $v - \beta \in C$ since $C$ is O-symmetric. Therefore $v = \frac{(\beta+v)+(v-\beta)}{2} \in C$, which contradicts with assumption $v \notin C$. So without loss of generality assume that $\beta + v \notin C$. Note that without loss of generality we can also assume that $x_i \in D'$ with good probability. Now, we apply the argument as explained in Regev's notes [Re] to reason with a modified distribution of the $x_i$. As $x_i \in D'$ we can replace $x_i$ by $x_i - v$. So it is easy to see that after sieving with good probability there exists tuples $(x_i, z_i), (x_j, z_j) \in S$ such that $r_{i,j} = (z_i - x_i) - (z_j - x_j) = v + \beta_i - \beta_j$. So $r_{i,j} = v + \beta \notin C$ and clearly $\|r_{i,j}\|_p \leq (1+\epsilon)\|v\|_p$ since $\|\beta_i - \beta_j\|_p \leq \epsilon$. It is easy to see that the algorithm runs in time $2^{O(n \log(1/\epsilon))} poly(size(\mathcal{L}))$. This completes the proof of the theorem. ∎

Next we give an algorithm to solve the Theta-Series problem using Theorem 3.

**Corollary 4.** *Given an integer lattice $\mathcal{L} \subseteq \mathbb{Z}^n$ of rank $n$ and an integer $k$, there is a $2^{O(n \log(kp))} \cdot poly(size(\mathcal{L}))$ time randomized algorithm to decide whether $\mathcal{L}$ contains a vector $v$ such that $\|v\|_p^p = k$.*

*Proof.* First we describe the algorithm. Let $m \in \mathbb{Q}$ such that $(k-1)^{1/p} < m \leq (k-1/2)^{1/p}$. Let $C$ be the convex body $B_p(0, m)$. We choose $\epsilon < (1 + \frac{1}{2k})^{1/p} - 1$ and run the $1 + \epsilon$ approximation algorithm described in the Theorem 3 on the instance $(\mathcal{L}, C)$ of CAP with respect to $\ell_p$ norm. Output "YES" if the algorithm outputs a vector $u$ such that $\|u\|_p^p = k$ otherwise output "NO". To see the correctness of the algorithm, suppose that $\mathcal{L}$ contains a vector $v$ such that $\|v\|_p^p = k$. It is easy to see that $v$ is shortest vector in $\mathcal{L} \setminus C$. The approximation algorithm in Theorem 3 outputs a vector $u \in \mathcal{L} \setminus C$ such that with good probability $\|u\|_p \leq (1+\epsilon) \cdot \|v\|_p$. This implies $\|u\|_p^p < (1 + 1/2k) \cdot k = k + 1/2$. But since $\mathcal{L}$ is an integer lattice this in fact implies $\|u\|_p^p = k$. It is easy to see that the algorithm runs in time $2^{O(n \log(kp))} \cdot poly(size(\mathcal{L}))$. ∎

# 6 A lower bound result

The AKS sieving procedure is a breakthrough technique for showing upper bounds for various lattice problems. In this paper too we have provided new examples. The AKS technique is actually generic; as we have shown in this paper, it works for all nice gauge functions. In this section we prove a lower bound result showing that essentially this generic property of AKS sieving makes it unlikely to give algorithms that are asymptotically faster than the $2^{O(n)}$ bound.

Consider the problem of finding *all* shortest nonzero vectors in a lattice $\mathcal{L} \subset \mathbb{R}^n$ with respect to a nice gauge function $f : \mathbb{R}^n \to \mathbb{R}$. The algorithms we consider take a basis for $\mathcal{L}$ as input and the nice gauge function $f$ is given by oracle access: the oracle outputs $f(x)$ for a query $x$. The *query complexity* of the algorithm is the worst-case number of queries that it makes for inputs $(\mathcal{L}, f)$. Now, the AKS sampling algorithm [AKS01,Re] can be used to compute all the shortest vectors in the integer lattice $\mathcal{L}$ with respect to $\ell_2$ norm in time $2^{O(n)}$. As explained in this paper (in Section 3 and later), the AKS-sampling can be easily adapted to work for any *nice* gauge function $f$ where $f$ is given by oracle access.

We now show a lower bound on the query complexity of any such algorithm based on an adversary argument. For the lower bound theorem we fix the lattice as the standard lattice $\mathbb{Z}^n$ and consider only those nice gauge functions $f : \mathbb{R}^n \to \mathbb{R}$ such that there are at most $n^2$ many shortest nonzero vectors in $\mathbb{Z}^n$ with respect to $f$. Let $\mathcal{F}$ denote this family of nice gauge functions. The proof of the following theorem is given in the appendix.

**Theorem 4.** *The query complexity of any deterministic/randomized algorithm that takes as input the standard lattice $\mathbb{Z}^n$ and a nice gauge function $f \in \mathcal{F}$ as an oracle and outputs a list of all shortest nonzero vectors in $\mathbb{Z}^n$ with respect to the gauge function $f$ is $2^{\Omega(n)}$.*

# 7 A natural parameterization for lattice problems

In this section we introduce and study a natural parameterized version for lattice problems, motivated by similar problems for codes that have been studied in the parameterized setting by Downey et al [DFVW99,DF99].

Let $\mathcal{L} \subset \mathbb{R}^n$ be an integer lattice given by a basis $b_1, b_2, \cdots, b_n$. The *support* of a lattice point $\sum_{i=1}^n \alpha_i b_i$ w.r.t. the given basis $b_1, b_2, \cdots, b_n$ is the number of nonzero $\alpha_i$. We now define the *parameterized feasible set* of lattice points for parameter $k$ and w.r.t. the given basis $b_1, b_2, \cdots, b_n$ to be

$$\mathcal{F}_k(\mathcal{L}) = \{\sum_{i=1}^n \alpha_i b_i \in \mathcal{L} \mid \sum_{i=1}^n \alpha_i b_i \text{ has support at most } k\}.$$

This immediately gives a natural parameterization for all the standard optimization problems for lattices. The problems pSVP, pCVP, and pSAP are the parameterized versions of the shortest vector problem, closest vector problem, and the subspace avoiding problem respectively, where the input instances come with a parameter value $k$ and the feasible set for these problems is $\mathcal{F}_k(\mathcal{L})$, where $\mathcal{L}$ is the input lattice with a given basis.

Based on the AKS sampling procedure we can easily show that the problems pSVP, pCVP, and pSAP all have randomized algorithms with running time $(n + k)^{O(k)}$. On the other hand, we show that pCVP and pSAP are hard for W[1] w.r.t. fpt reductions.

**Theorem 5.** *The parameterized problems* pSVP, pCVP, *and* pSAP *all have randomized algorithms with time bound $(n + k)^{O(k)} \cdot \text{poly}(s)$, where $s$ is the input size encoded in binary.*

*Proof.* We first explain the simple algorithm for pCVP. The problem pSVP is similarly solved, and pSAP can be solved by using the algorithm for pCVP. Let $(v, \mathcal{L}, k)$ be input instance for pCVP, where $\mathcal{L}$ is given by basis $b_1, b_2, \cdots, b_n$. We will generate $\binom{n}{k}$ many lattices $\mathcal{L}_S, S \subset [n], |S| = k$, where $\mathcal{L}_S$ is the rank $k$ lattice generated by the $k$ vectors $\{b_j \mid j \in S\}$. The feasible set $\mathcal{F}_k$ is the union of all these lattices $\mathcal{L}_S$. Thus, it suffices to solve each of the $\binom{n}{k}$ CVP instances $(v, \mathcal{L}_S)$ and pick the best solution among them. By using Kannan's algorithm [Kan87] we can solve each of these instances in time $2^{k \log k} \cdot \text{poly}(\text{size}(v, \mathcal{L}))$. Thus, we can solve the pCVP instance in time $n^k \cdot 2^{k \log k} \cdot \text{poly}(\text{size}(v, \mathcal{L}))$, as claimed. The algorithm for pSVP is exactly on the same lines. For pSAP we need to apply ideas from Theorem 2. We showed in Theorem 2 that for an instance $(\mathcal{L}, M)$ of SAP, where $M$ is $\ell$-dimensional, we can find the shortest vector in $\mathcal{L} \setminus M$ by solving $2^{O(n)}$ instances of the kind $(v_i, \mathcal{L} \cap M)$ and taking the best solution among them. The key point here is that $\mathcal{L} \cap M$ is a rank $\ell$ lattice implying that Kannan's algorithm finds exact solution in time $2^{\ell \log \ell} \cdot \text{poly}(s)$, where $s$ is a bound on input size. Applying the same idea to pSAP, we will first generate the $\binom{n}{k}$ many instances $(\mathcal{L}_S, M)$ of instance of SAP, where $M$ is $\ell$-dimensional and $\mathcal{L}_S$ has rank $k$, where $k$ is the parameter. Note that we can easily obtain lattice $L'_S$ in a $k$ dimensional space $R$ by applying suitable linear transformation. Let $M' = M \cap R$. It is clear that $dim(M') = t \leq k$. Now, applying the method of Theorem 2 we can exactly solve of the SAP instances $(L'_S, M')$ by solving $2^{O(k)}$ many instances of rank $t$ CVP instances, where $t \leq k$. The overall running time is easily seen to be $(n+k)^{O(k)} \cdot \text{poly}(s)$. ∎

We next show that pCVP is hard for W[1] (we show it for $\ell_2$ norm which can be easily extended to any $\ell_p$ norm). We leave the question open whether pSVP is W[1]-hard or is fixed parameter tractable.

**Theorem 6.** *The parameterized problem* pCVP *is hard for* W[1] *under fpt many-one reductions.*

*Proof.* We give an fpt many-one reduction from the $k$-perfect code problem to pCVP. $k$-perfect code problem is defined in [DFVW99] where it is shown to be W[1]-hard. An input instance of $k$-perfect code is $(G, k)$, where $G = (V, E)$ is an undirected graph and the problem is whether there is a subset $S \subset V$ such that $|S| = k$ and $N(x), x \in S$ is a *partition* of $V$, where $N(x)$ is a neighbourhood of $x$. Let $|V| = n$. We can represent $N(x)$ for $x \in V$ as an $n$-dimensional 0-1 vector $v_x$ that has a 1 in the $i^{th}$ position if and only if $i \in N(x)$. Thus, given a collection of these vectors $v_i, i \in V$, we are asking if some $k$-subset of the $v_i$'s adds up to the all 1's vector. Let $w_i, i \in V$ be $n$-dimensional vectors, where each $w_i$ has a large positive integer $M$ in the $i^{th}$ position and is 0 elsewhere. To continue the reduction, for each $i \in V$ we replace $v_i$ with a $2n$-dimensional vector $u_i$ whose first $n$ coordinates is the vector $v_i$, the next $n$ coordinates is the vector $w_i$. Finally, we consider the vector $u$ whose first $n$ coordinates are all 1's, the next $n$ coordinates are all $M$'s. Let $\mathcal{L} \subset \mathbb{R}^{2n}$ denote the integer lattice with basis $B = \{u_i \mid i \in V\}$, and consider the pCVP instance $(u, \mathcal{L}, k)$. We claim that there is a vector $v \in \mathcal{L}$ that is in $\mathcal{F}_k(\mathcal{L})$ such that $||v - u||_2 \leq (n-k)M$ if and only if $(G, k)$ is a yes instance of the $k$-perfect code problem. If $(G, k)$ is a yes instance then clearly there is such a vector $v \in \mathcal{F}_k(\mathcal{L})$. For the converse, suppose $v \in \mathcal{F}_k(\mathcal{L})$ such that $||v - u||_2 \leq (n-k)M$. Suppose $v = \sum_{j=1}^{k} \alpha_{i_j} u_{i_j}$. Notice that in any case $||v - u||_2 \geq (n-k)M$. Furthermore, the last inequality is strict unless all the $\alpha_{i_j} = 1$. This in turn forces that $(G, k)$ is a yes instance of the $k$-perfect code problem. This completes the proof. ∎

## 8 The complexity of unique-CVP

In this section we consider the complexity of the search version of promise problem unique-CVP. Given an integer lattice $\mathcal{L} \subseteq \mathbb{Q}^n$ and $v \in \mathbb{Q}^n$ with the promise that the closest vector to $v$ in $\mathcal{L}$ is unique then the problem unique-CVP is to find such a vector. We show that the search version of unique-CVP with respect to any $\ell_p$ norm, $1 \leq p < \infty$ is as hard as search version of CVP. The proof is by a randomized many-one reduction of the search version of CVP to search version of unique-CVP. We can also show a similar reduction from search version of SVP to search version of unique-SVP for any $\ell_p$ norm. The unique-SVP problem has gained importance after Ajtai et al in [AD97] proposed a public-key cryptosystem whose security depends upon the worst-case hardness of a variant of unique-SVP. R. Kumar and D. Sivakumar in [KS99] have shown that the decision version of unique-SVP with respect to $\ell_2$ norm is NP-hard under randomized reductions. More precisely, they show for a given lattice $\mathcal{L}$ and number $d$, checking if there is a $v \in \mathcal{L}$ such that $||v||_2 \leq d$ is NP-hard under randomized reductions, given that $\mathcal{L}$ has at most one vector (upto the sign) of length at most $d$. Their reduction is similar to the Valiant-Vazirani reduction for uniqueSAT [VV85].

Our reduction for unique-CVP is based on a general form of the isolation lemma [MVV87] due to Klivans-Spielman [KS01] in which random weights are assigned to "isolate" a linear form from a collection. We give a brief sketch of the reduction and its correctness argument. By scaling up, we can assume that the CVP instance $(\mathcal{L}, v)$ is such that the input lattice $\mathcal{L} \subseteq \mathbb{Z}^n$ and $v = (v_1, \cdots, v_n) \in \mathbb{Z}^n$. We will make a suitable random scaling of coordinates to obtain a new instance $(\mathcal{M}, v')$ of CVP such that there is unique closest vector to $v'$ in $\mathcal{M}$ with high probability, and moreover we can easily recover a closest vector to $v$ in $\mathcal{L}$ given a closest vector to $v'$ in $\mathcal{M}$. We describe the reduction for the $\ell_2$ norm. The same reduction will essentially work for all $\ell_p$ norms.

Let $b_1, b_2 \cdots, b_n \in \mathbb{Z}^n$ be a basis of $\mathcal{L}$. Since we can replace $v$ by $v(\bmod \mathcal{L})$ which is easy to compute, there is no loss of generality in assuming $v$ lies in the fundamental parallelepiped of $\mathcal{L}$. The following claim is easy to show.

*Claim.* Let $\mathcal{L} \subseteq \mathbb{Z}^n$ be a lattice and $v \in \mathbb{Z}^n$. If $N$ is equal to number of closest vector to $v$ in $\mathcal{L}$ then $size(N) \leq (size(\mathcal{L}, v))^c$, for some absolute constant $c > 0$.

For the reduction, we pick $a_1, a_2, \cdots, a_n$ uniformly at random from $\{0, 1, \cdots, p\}$ where $p = 8n \cdot (size(\mathcal{L}, v))^{2c}$. Clearly, by the above claim $p \geq 8nN^2$. Let $K \in \mathbb{Z}^+$ that we will choose suitably large. Define $f_K : \mathbb{R}^n \longrightarrow \mathbb{R}^n$ as $f_K(x_1, \cdot, x_n) = ((K + a_1)x_1, \cdots, (K + a_n)x_n)$. Clearly, $\mathcal{M}_K = \{f_K(x) | x \in \mathcal{L}\}$ is an integer lattice with basis $\{f_K(b_1), f_K(b_2), \cdots, f_K(b_n)\}$. The following claim is easy to prove by direct calculation and bounding of $\ell_2$ norms.

*Claim.* Let $s = \text{size}(\mathcal{L}, v)$. Then there is an absolute constant $c' > 0$ such that for any $K \geq 2^{s^{c'}}$, if $f_K(x) \in \mathcal{M}$ is a vector closest to the vector $Kv$, then $x \in \mathcal{L}$ is a closest vector to $v$.

From the above claim it follows that given a closest vector to $Kv$ in $\mathcal{M}$ we can compute a closest vector to $v$ in $\mathcal{L}$ in polynomial time. Next we prove that $Kv$ has a unique closest vector in the lattice $\mathcal{M}$ with good probability.

Let $C = \{x \in \mathcal{L} | x \text{ is a closest vector to } v\}$. By the previous claim it is clear that for any vector $z \in \mathcal{M}$ closest to $Kv$ there is $x \in C$ such that $z = f_K(x)$. For $x = (x_1, \cdots x_n) \in \mathbb{R}^n$ let $wt(x) = \Sigma_{i=1}^n ((K + a_i)x_i - Kv_i)^2$. An index $i, 1 \leq i \leq n$ is a *bad index* if there are $x = (x_1, \cdots, x_n), y = (y_1, \cdots, y_n) \in C$ such that $wt(x) = wt(y)$ and $x_i \neq y_i$.

**Lemma 2.** *An index $i$, $1 \leq i \leq n$ is a* bad index *with probability at most $1/2n$, where the probability is over the random choice of $a_1, \cdots, a_n$. Hence there is a bad index with probability at most $1/2$.*

*Proof.* Recall that we pick each $a_i$ independently and uniformly at random from $0 \leq a_i \leq 8ns^{2c}$. Assume we have randomly picked all $a_j$ for $j \neq i$, where $i$ is some specific index. Since all $a_j, j \neq i$ are already picked, for every $x \in C$ notice that $wt(x) = ((K + a_i)x_i − Kv_i)^2 + N_x$ is a quadratic polynomial in the indetermintes $a_i$, where $N_x$ is a fixed integer that depends only on $x$ and $v$. Thus, we have an associated quadratic for each $x \in C$, and as already argued $|C| \leq s^c$. Now, by Bezout's theorem any 2 quadratic curves can intersect in at most four points. Thus, if we consider pairwise intersection of the quadratics we have at most $4s^{2c}$ distinct intersection points. Clearly, if $a_i$ takes a value different from any of these intersection points then $wt(x) \neq wt(y)$ for distinct $x, y \in C$. Since $a_i$ is uniformly picked from $[0..8ns^{2c}]$, the index $i$ is a bad index with probability at most $1/2n$. As a consequence of the union bound it follows that there is a bad index with probability at most $1/2$. This proves the lemma. ∎

Clearly, if there is no bad index for a random assignment to the $a_i$, there is a unique closest vector in $\mathcal{M}$ to $Kv$. Hence the randomized many-one reduction succeeds with probability at least $1/2$. This completes the correctness proof. It is easy to see that above argument can be applied for any $\ell_p$ metric. We summarize the result below.

**Theorem 7.** *For any $\ell_p$ norm there is a randomized polynomial time many-one reduction from the search version of CVP to the search version of unique-CVP.*

With some modifications to the randomized reduction, we can show that the search version of SVP (and SAP) is reducible to search version of unique-SVP (respectively unique-SAP).

## A  Proof of Theorem 4

*Proof Sketch.* We sketch the adversay argument only for deterministic algorithms (the randomized case is essentially the same). Notice that the $\ell_2$ norm is in $\mathcal{F}$ as there are exactly $2n$ shortest nonzero vectors in $\mathbb{Z}^n$ w.r.t. $\ell_2$ norm. This is the set $U = \{\pm e_1, \pm e_2, \cdots, \pm e_n\}$ where $e_i$ are the standard basis vectors in $\mathbb{Z}^n$.

Now, suppose $\mathcal{A}$ is an algorithm that makes at most $q(n) = o(2^n)$ queries on any input. The adversary will answer all queries w.r.t. the $\ell_2$ norm. Let $T = \{x_1, x_2, \cdots, x_k\}, k \leq q(n)$ be the set of vectors queried by $\mathcal{A}$. We can assume without loss of generality that $U \subseteq T$. Specifically, the adversary outputs $f(x_i) = \|x_i\|_2 = a_i$ for each query $x_i \in T$. Notice, for each query $v_i \in U$, that the adversary outputs $f(v_i) = 1 = \|v_i\|_2$.

Finally, suppose the algorithm outputs a set $S$ as the set of shortest vectors. By choice of $\mathcal{F}$, $|S| \leq n^2$. If $U \not\subseteq S$ then the adversary can set the actual input gauge function to $\ell_2$ implying that the algorithm $\mathcal{A}$ is incorrect. Thus, we can assume $U \subseteq S$.

Consider the $2^n$ quadrants of $\mathbb{R}^n$ formed by the standard coordinate axes. We can describe the $2^n$ quadrants $Q_z$ by vectors $z \in \{1, -1\}^n$, where the quadrant $Q_z$ consists of all points $(x_1, x_2, \cdots, x_n) \in \mathbb{R}^n$, where $x_i$'s are all nonzero and $x_i$ and $z_i$ have the same sign. Notice that each such $z \in \mathbb{Z}^n$ is the unique lattice point in quadrant $Q_z$ with $\ell_2$ norm $\sqrt{n}$. Since $|T \cup S| = o(2^n)$, for sufficiently large $n$ we have $|T \cup S| < 2^{n-1}$. Hence there exists a lattice point $y = (y_1, \cdots, y_n) \in \{1, -1\}^n$ such that no point from $T \cup S$ lies in the quadrants $Q_y$ or $Q_{-y}$.

Next we define an O-symmetric convex body $C$. Consider the quadrants $Q_y$ and $Q_{-y}$.

The bounding axes of quadrant $Q_y$ intersect the sphere $B_2(0, 1)$ in the points $P_i = (0, 0, \cdots, 0, y_i, 0, \cdots, 0)$ for $i = 1$ to $n$. Clearly, $P_1, \cdots, P_n$ lie on a $n-1$ dimensional hyperplane that intersects $B_2(0, 1)$ in an $n-1$ dimensional sphere $M_y$. Let $C_y$ be the $n$-dimensional right-circular cone with base $M_y$ and vertex of cone at $y$. Similarly, we obtain a right circular cone $C_{-y}$ corresponding to $-y$. We define $C = C_y \cup C_{-y} \cup B_2(0, 1)$. It is easy to see that $C$ is an O-symmetric convex body. Let $T' = \{\frac{v}{\|v\|_2} \mid v \in T\}$. Clearly, the points $\{y, -y\} \cup T'$ lie on the surface of convex body $C$. By Proposition 1 this defines a gauge function $f$ such that $B_f(0, 1) = C$. Notice that $f$ is a nice gauge function since $B_2(0, 1) \subseteq C \subseteq B_2(0, \sqrt{n})$. Hence we have a nice gauge function $f$ and a lattice point $y \in \mathbb{Z}^n \setminus S$ such that $f(y) = 1$ and for all $x'_i \in T'$, $f(x'_i) = 1$. Thus, the shortest vector set for this gauge function is the set $U \cup \{y, -y\}$ which is not the set output by the algorithm $\mathcal{A}$. This proves the theorem. ∎